\documentclass[fleqn,10pt]{wlscirep}
\usepackage[utf8]{inputenc}
\usepackage[T1]{fontenc}
\usepackage{graphicx}
\usepackage{geometry}
\usepackage{amsmath,amsthm,amssymb,amsfonts}
\usepackage{ulem}  
\usepackage{fancybox,mathrsfs,version,bm}
\usepackage{epsfig,graphicx}
\usepackage{url,hyperref}
\usepackage{color,xcolor}
\usepackage{tabularx,booktabs,multirow}
\usepackage{xstring}
\usepackage{subfig}
\usepackage{mathtools}
\usepackage{makecell}
\usepackage{tikz-cd}
\usepackage{ulem} 
\usepackage{caption}  
\usepackage{times}
\usepackage{cite}
\usepackage{multicol}
\usepackage{lineno}

\title{The Gestalt Computational Model by Persistent Homology}
\author[1] {Yu Chen} 
\author[1,*] {Hongwei Lin}
\author[1] {Jiacong Yan}
\affil[1] {School of Mathematics Science, Zhejiang University,No. 866, Yuhangtang Road, Hangzhou, China}
\affil[*] {hwlin@zju.edu.cn}

\begin{abstract}
   Widely employed in cognitive psychology, 
   		Gestalt theory elucidates basic principles in visual perception.
   	However, the Gestalt principles are validated mainly by psychological experiments,
   		lacking quantitative research supports and theoretical coherence. 
   In this paper, we utilize persistent homology, 
   		a mathematical tool in computational topology, 
   		to develop a unified computational model for Gestalt principles, 
   		addressing the challenges of quantification and computation. 
   On the one hand, the Gestalt computational model presents quantitative supports for Gestalt theory.
   		On the other hand, it shows that the Gestalt principles can be uniformly calculated using persistent homology, 
   		thus developing a coherent theory for Gestalt principles in computation. 
   Moreover, it is anticipated that the Gestalt computational model can serve as a significant 	
   		computational model in the field of computational psychology, 
   		and help the understanding of human being visual perception.
   
\end{abstract}

\begin{document}	

\flushbottom
\maketitle
\thispagestyle{empty}
%\begin{multicols}{2}

Formulated by psychologists Wolfgang Köhler, Kurt Koffka, and Max Wertheimer in the early 20th century, 
the Gestalt theory has evolved over time, offering a classic framework for understanding human visual perception \cite{kohler1967gestalt,koffka2013principles,wertheimer1923laws}.
Researches on Gestalt theory have demonstrated the extraction of global properties in visual perception primarily through experiments. \cite{chen1982topological, zhuo2003contributions, chen2005topological, wolfe2004attributes, han1999uniform, he2009connectedness}. 
As cognitive psychology continues to evolve, 
	the importance of establishing computational models becomes increasingly prominent, 
	which can offer a detailed, process-oriented approach that goes beyond mere experimental validation \cite{sun2008cambridge}. 
These models provide algorithmic specificity and precision, allowing for a deeper understanding of cognitive processes \cite{sun2008cambridge}. 
Furthermore, computational cognitive modeling serves as a potent tool for theory building and hypothesis generation, 
	allowing researchers to explore the intricacies of the human mind in a manner that traditional experimentation cannot fully achieve \cite{sun2008cambridge}.
In this study, we aim to develop a computational model of the Gestalt theory.
Specifically, we utilize persistent homology, 
	an essential tool in computational topology, 
	to compute visual perceptual results that conform to Gestalt theory. 
On the one hand, this validates the extraction of global properties in the visual perception system by computation, 
 	rather than relying solely on experiments. 
On the other hand, 
 	given that our computational model can efficiently compute and extract the global topological features described by Gestalt theory from perceived objects, it has the potential to provide new insights into Gestalt theory and to serve as a significant computational model for visual perception in the field of computational psychology \cite{sun2008cambridge}.

Emphasizing on organizing sensory information into coherent patterns and wholes, 
 	Gestalt theory asserts that each component of any visual perceptual outcome is interrelated, 
 	and the entirety is shaped by these connections. 
In essence, Gestalt theory explores the relationship between the whole and its parts, 
 	with the fundamental premise that the visual object is initially perceived as a unified whole and subsequently as parts. 
One prominent theory that describes this relationship is visual topology theory \cite{chen1982topological}, which is developed to clarify the global properties of Gestalt theory in terms of topological concepts.
Visual topology theory interprets the concept of whole in Gestalt theory as large-scale topological features intrinsic 
	to the perceived object, 
 	and then explores the relationship between the global and the local aspects in Gestalt theory from the perspective of algebraic topology \cite{chen2005topological}.
 	This theory successfully establishes a mathematical model of Gestalt theory, 
 	but it is incomputable.

Actually, quantifying and computing Gestalt theory has always been a significant challenge.
Several attempts have been made to quantify Gestalt theory \cite{jakel2016overview}. 
 Statistical methods are commonly utilized to quantify Gestalt principles. For instance, the Helmholtz principle and the Number of False Alarms are employed to detect specific structures in an image that reflect Gestalt principles \cite{desolneux2007gestalt}, which is a typical statistical method. Another common method, Bayesian hierarchical grouping, also based on statistics, addresses Gestalt organization as a Bayesian statistical inference problem \cite{froyen2015bayesian}. However, Bayesian hierarchical grouping primarily focuses on the Gestalt principles associated with grouping problems. And these statistical methods may require adjustment of numerous parameters and may lead to complex computations.
 Furthermore, Hawkins et al. \cite{hawkins2016can} proposed capacity coefficients to quantify the gap between the whole 
 	and the sum of its parts. 
 Wei et al. \cite{wei2018objective} introduced a method using the tilt aftereffect from visual adaptation to quantify grouping effects. 
 Chen et al. \cite{chen1982topological,chen2005topological} quantified the Gestalt theory using the homology theory within tolerance space 
 	to interpret Gestalt theory from the perspective of visual topology theory. 
 	But these studies only involved classical psychological experiments and did not provide more direct computational models for the quantitative analysis of Gestalt principles.
 Peng et al. \cite{peng2021computational} also employed tolerance space to compute the proximity and similarity principle 
 	for dot-pattern grouping. 
 But this approach only addresses these two principles and is not generalized to some other key Gestalt principles 
 	such as closure and pragnanz. 
 In conclusion, the Gestalt principles are quntified in different ways, 
 	lacking a coherent computational model. 
 
 Persistent homology, a novel computational tool that makes the classical homology theory computable, 
 	currently serves as a method for identifying topological features within target shapes \cite{edelsbrunner2022computational,zomorodian2004computing,edelsbrunner2002topological,edelsbrunner2008persistent,bubenik2007statistical,cohen2005stability}. 
 The great significance of persistent homology lies in making many abstract concepts in algebraic topology 
 	efficiently computable, 
 	thus establishing a new branch known as computational topology. 
 Moreover, topological data analysis (TDA), 
 	a powerful new branch for data processing, 
 	has been developed based on methods including persistent homology \cite{zomorodian2012topological,wasserman2018topological,munch2017user,chazal2021introduction}. 
 It has been successfully applied in various fields such as biomedicine \cite{skaf2022topological}, oncology \cite{bukkuri2021applications}, 
 	chemical engineering \cite{smith2021topological}, and machine learning \cite{carlsson2012topological,khasawneh2018chatter,muszynski2019topological,townsend2020representation,meng2021persistent,pun2022persistent}, etc. 

 Since visual topology theory utilizes classical algebraic topology to interpret Gestalt theory, 
  	and persistent homology makes algebraic topology concepts efficiently computable,  
  	persistent homology can be employed to calculate key Gestalt principles, 
  	e.g., similarity, proximity, closure, good continuation and pragnanz \cite{koffka2013principles,wertheimer1923laws,goldstein1989sensation}, in a coherent way. 
  In this paper, we will elaborate on the underlying mechanisms of these principles within the context of 
  	persistent homology, 
  	and develop the coherent computational model for Gestalt theory. 
  The computation results demonstrate that persistent homology is an efficient and straightforward tool
  	 for calculating Gestalt theory.

\section*{Results}

\subsection*{Gestalt computational model}
 We now introduce the computational model for Gestalt theory, using persistent homology and the corresponding persistence diagram (PD) based on the Vietoris--Rips complex (VR complex) and Vietoris--Rips filtration (VR filtration) \cite{edelsbrunner2022computational}. 
 It should be noted that, 
 in the VR filtration, 
 	a series of nested VR complex will be generated
 	$$ VR(\varepsilon_0) \subset VR(\varepsilon_1) \subset \cdots \subset VR(\varepsilon_i)\subset VR(\varepsilon_{i+1}) \subset \cdots \subset VR(\varepsilon_N), $$
 where 
 $\varepsilon_0=0 < \varepsilon_1 < \cdots<\varepsilon_i<\varepsilon_{i+1}<\cdots< \varepsilon_N$. 
 Moreover, in the PDs presented in this paper, 
 	the red points represent points in the zero-dimensional persistence diagram, 
 	and the blue points represent points in the one-dimensional persistence diagram.  
 The computational model for Gestalt theory is detailed as follows, 
 	where the perceived objects are represented as a planar point set 
\begin{equation}  
	\label{eq:pt_set}  
	\{\bm{P}_i = (x_i, y_i), i=1,2,\cdots,n\}.  
\end{equation} 
 Here for an object that appears to have a size, 
 	such as a dot, $(x_i, y_i)$ represents its barycenter coordinate.
\begin{itemize}
 	\item [1.] {\it Extra coordinates assignment:} Assign each point extra $z$-coordinates according to the attributes influencing perception, 
 		that is, color, shape, size, to name a few. 
 		And the planar point set is changed to the point set in $(m+2)$-dimensional space, i.e., 
 		\begin{equation} \label{eq:extra_coord}
 			\{\bm{Q}_i = (x_i, y_i, z_{i,1}, z_{i,2}, \cdots, z_{i,m})\}
 		\end{equation}
 	where $i=1,2,\cdots,n$.
 	The extra $z$-coordinates are taken according to features, 
 		e.g., color, shape, size, etc. 
 	\item [2.] {\it PD calculation:} Construct VR filtration based on the point set and compute the corresponding persistence
 		 diagrams. 
 		 Typically, only the zero-dimensional persistence diagrams (0-PDs) or one-dimensional persistence diagrams (1-PDs) is employed in the Gestalt computational model. 
 	\item [3.] {\it Point clustering:}  Project the points in a PD onto the line $y=-x$, 
 		and cluster the projected points into two classes using, for example,  $k$-means algorithm \cite{hamerly2002alternatives} ($k=2$).
 		While the class closer to the origin $(0,0)$ corresponds to the noise points, called \textit{noise class}; 
 			the class farther away to the origin corresponds to the significant points with greater persistence,
 			called \textit{significant class}. 
 	\item [4.] {\it Threshold determination:} Determine a suitable threshold $\varepsilon_g$.
 		For 0-PD, $\varepsilon_g$ should be greater than the largest death time $t_d$ of the noise points, 
 			and less than the smallest death time of the significant points; 
 		for 1-PD, $\varepsilon_g$ should be greater than the largest birth time $t_b$ of the significant points,
 			and less than the smallest death time of the significant points. 
 		At the VR complex with the parameter $\varepsilon_g$, i.e., $VR(\varepsilon_g)$, 
 			all significant topological features, such as significant connected components or loops, exist. 
 		By default, we set $\varepsilon = t_d$ (for 0-PD) or $\varepsilon = t_b$ (for 1-PD). 
 	\item [5.] {\it Perception result reconstruction:} Reconstruct the visual perceptual results from the VR complex with the parameter $\varepsilon_g$,
 		i.e., $VR(\varepsilon_g)$.  
 		The reconstructed visual perceptual results satisfy the Gestalt principles.  		
\end{itemize}

\begin{figure*}[ht]
	\centering
	\includegraphics[width=0.97\linewidth]{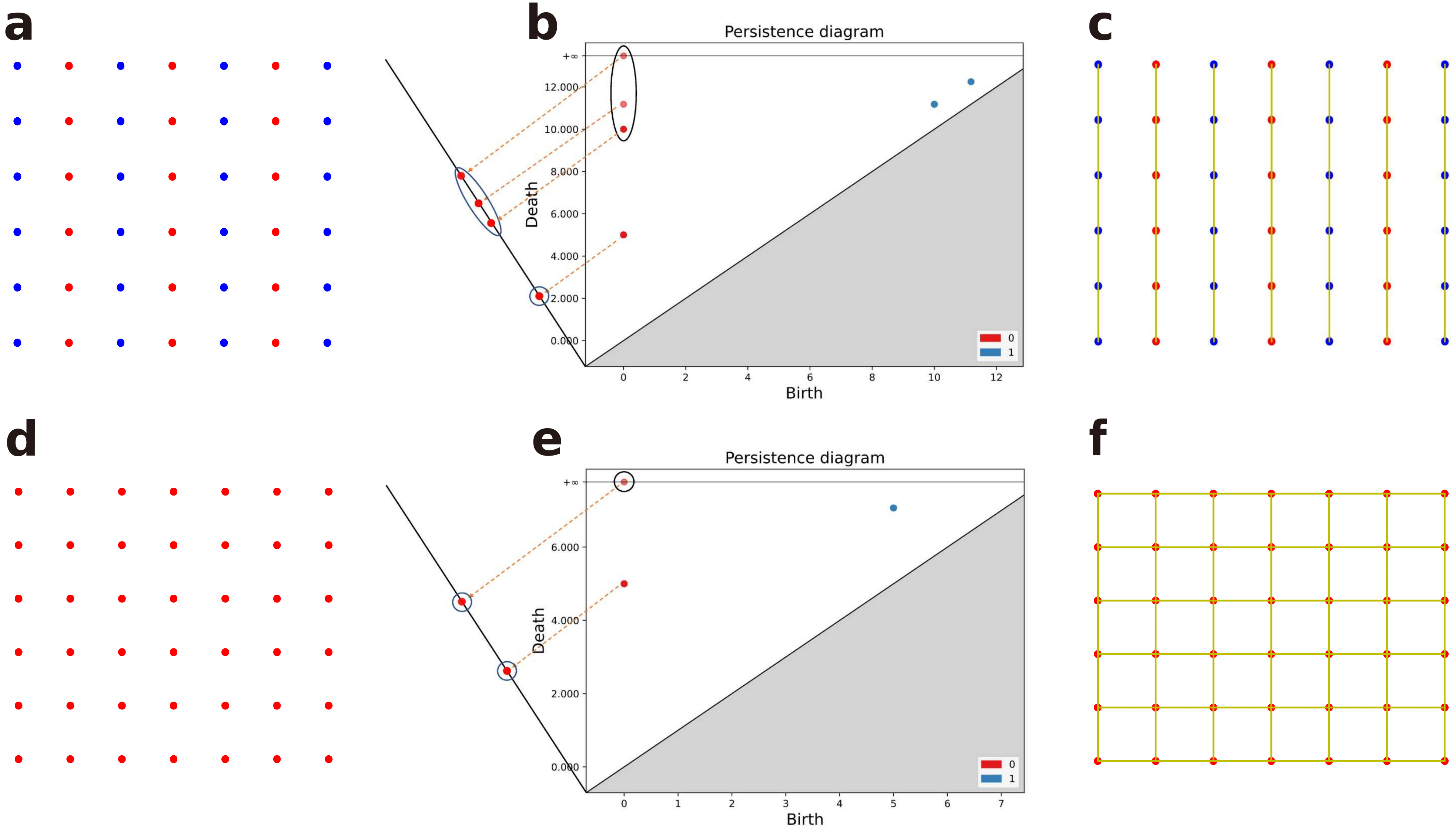}
	\caption{Calculation of Gestalt similarity principle.   
		a, d: Examples for similarity principle. 
		b, e: The clustering result of the corresponding 0-PD with highlighted zero-dimensional significant points. 
		c, f: The VR complex $VR(\varepsilon_g)$ that coincides with the similarity principle.}
	\label{fig:fig1}
\end{figure*}

 \begin{figure*}[ht]
	\centering
	\includegraphics[width=0.75\linewidth]{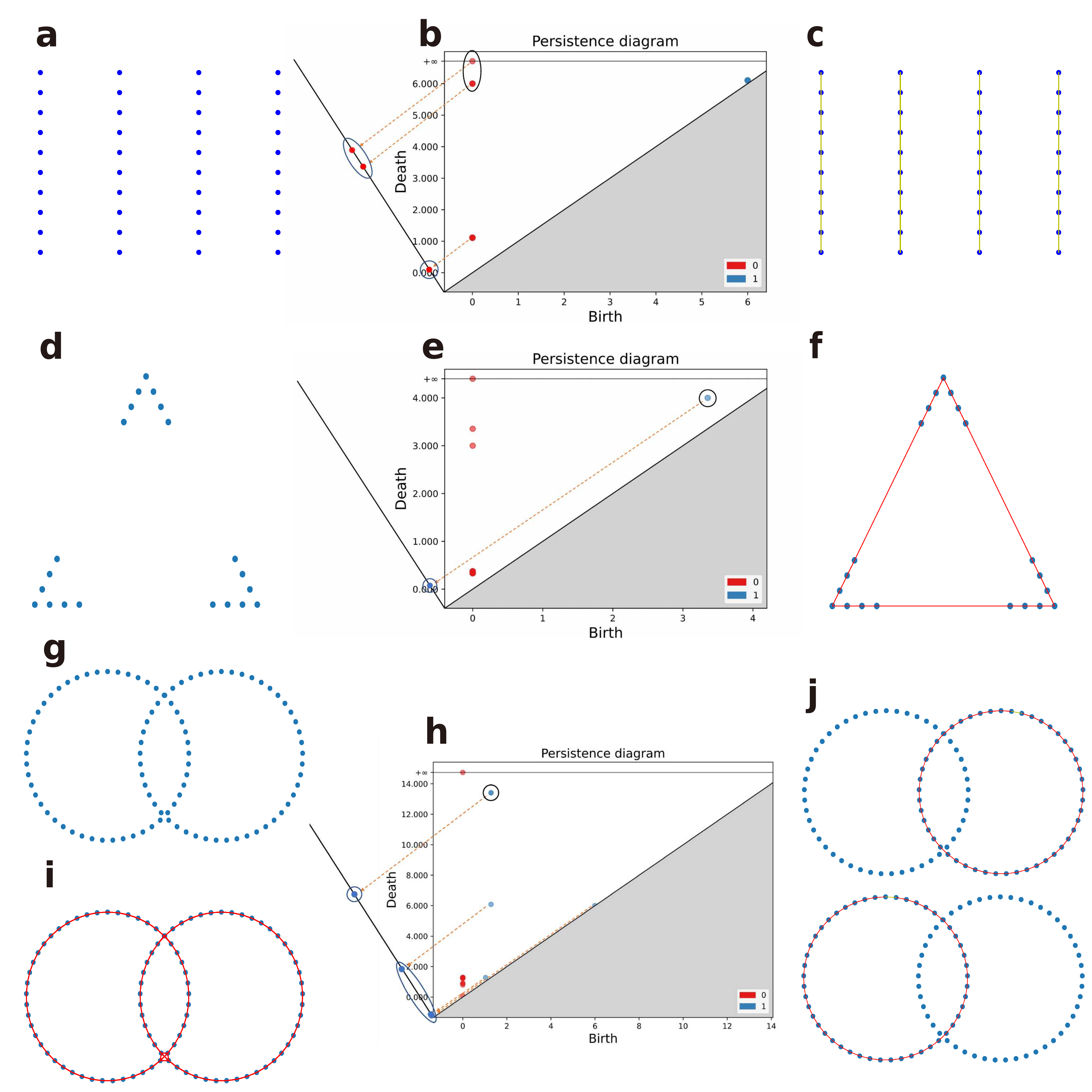}
	\caption{Calculation of the principle of proximity, closure and good continuation.  
		a, d, g: Examples of Gestalt proximity, closure and good continuation principles. 
		b: The clustering result of the corresponding 0-PD with highlighted zero-dimensional significant points. 
		c: The VR complex $VR(\varepsilon_g)$ that coincides with the Gestalt proximity principle.
		e, h: The clustering result of the corresponding 1-PD with highlighted one-dimensional significant points. 
		f: The hidden loop in the point cloud. 
		i: The corresponding 1-skeleton when all one-dimensional topological features (loops) have formed. 
		j: Two circles are successfully identified. Here edges whose two terminal nodes are the starting and ending points are drawn in yellow.}
	\label{fig:fig2}
\end{figure*}

\subsection*{Similarity principle}
 To demonstrate the Gestalt computational model, 
 we employ it to calculate the visual perceptual results dominated by the Gestalt similarity principle.
 According to the similarity principle, 
 when objects are relatively similar in shape, size, color or other attributes, 
 these objects appear to be grouped together. 
 In the following instance of calculating the similarity principle via the Gestalt computational model developed above,
 	only one extra coordinate $z_{i,1}$ for color is required  (see Fig. \ref{fig:fig1} a and d). 
 We set $z_{i,1} = 10$ for blue points and $z_{i,1}  = 0$ for red points. 
	When calculating the subsequent Gestalt principles, 
	$z_{i,1}$ is also set up in the same manner.
 First we take the point set in Fig. \ref{fig:fig1} a as an example. 
 After the extra coordinate for color is assigned, 
 	the VR filtration is constructed,
 	and the 0-PD is generated (refer to Fig. \ref{fig:fig1} b, and the construction of VR filtration is depicted in Supplementary Video 1 in the Supplementary Information).
 Next, the points in the 0-PD are projected onto the line $y=-x$, 
 	and they are clustered into two classes with 2-means clustering method.
 The significant class contains seven significant points with coordinates $(0,+\infty)$, $(0,11.18033981)$, 
 	and $(0,10)$ (occurring five times), 
 	and the noise class includes three points, each with coordinate $(0,5)$. 
 Consequently, we take the threshold $\varepsilon_g = 5$, 
 	and the VR complex $VR(5)$ (Fig. \ref{fig:fig1} c) is the visual perceptual result,
 	which satisfies the Gestalt similarity principle very well. 
 That is, the points with the same color are grouped together.  
 
 In Fig. \ref{fig:fig1} d, there is another example for similarity principle, i.e., a grid of red dots. 
 After constructing the VR filtration, 
 	generating the 0-PD, 
 	and clustering the points in the 0-PD (Fig. \ref{fig:fig1} e),
 	the threshold $\varepsilon_g$ can be set as $\varepsilon_g = 5$.
 The VR complex $VR(5)$ is demonstrated in Fig. \ref{fig:fig1} f, 
 	which is the calculation result by the Gestalt computational model. 
 The result in Fig. \ref{fig:fig1} f forms a lattice, 
 	satisfying the Gestalt theory perfectly. 
 
 In the following, we will briefly introduce the calculation of other Gestalt principles,
 	including proximity, closure, good continuation and pragnanz,
 	using the Gestalt computational model.
 Details of the calculation are provided in the Supplementary Information,
 	as well as the videos of the VR filtration procedure.

 \begin{figure*}[ht]
 	\centering 
 	\includegraphics[width=0.97\linewidth]{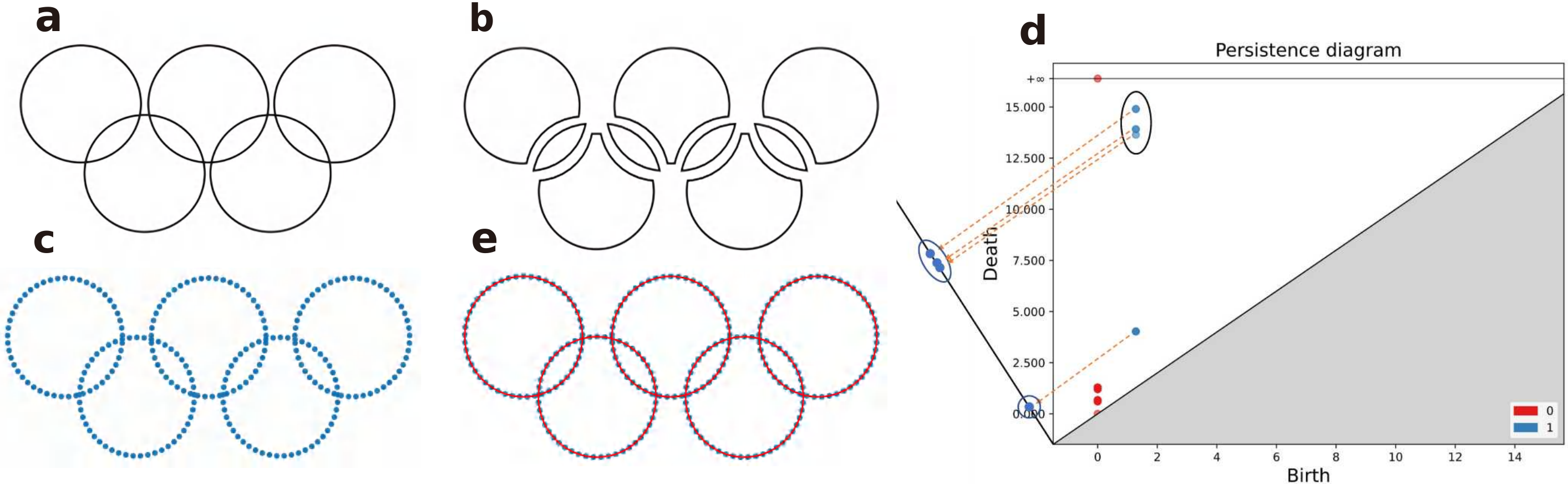}
 	\caption{Calculation of Gestalt pragnanz principle. 
 		a: The shape of the five Olympic rings. 
 		b: It is unlikely that the shapes in Fig. \ref{fig:fig3} a will be seen as nine parts. 
 		c: An example of the pragnanz principle. The point cloud represents the shape of the five Olympic rings. 
 		d: The clustering result of the corresponding 1-PD with highlighted one-dimensional significant points. 
 		e: Five circles that represent five significant topological features correspond to the five significant points in the 1-PD.}
 	\label{fig:fig3}
 \end{figure*}
 
 \begin{figure*}[ht]
 	\centering
 	\includegraphics[width=0.6\linewidth]{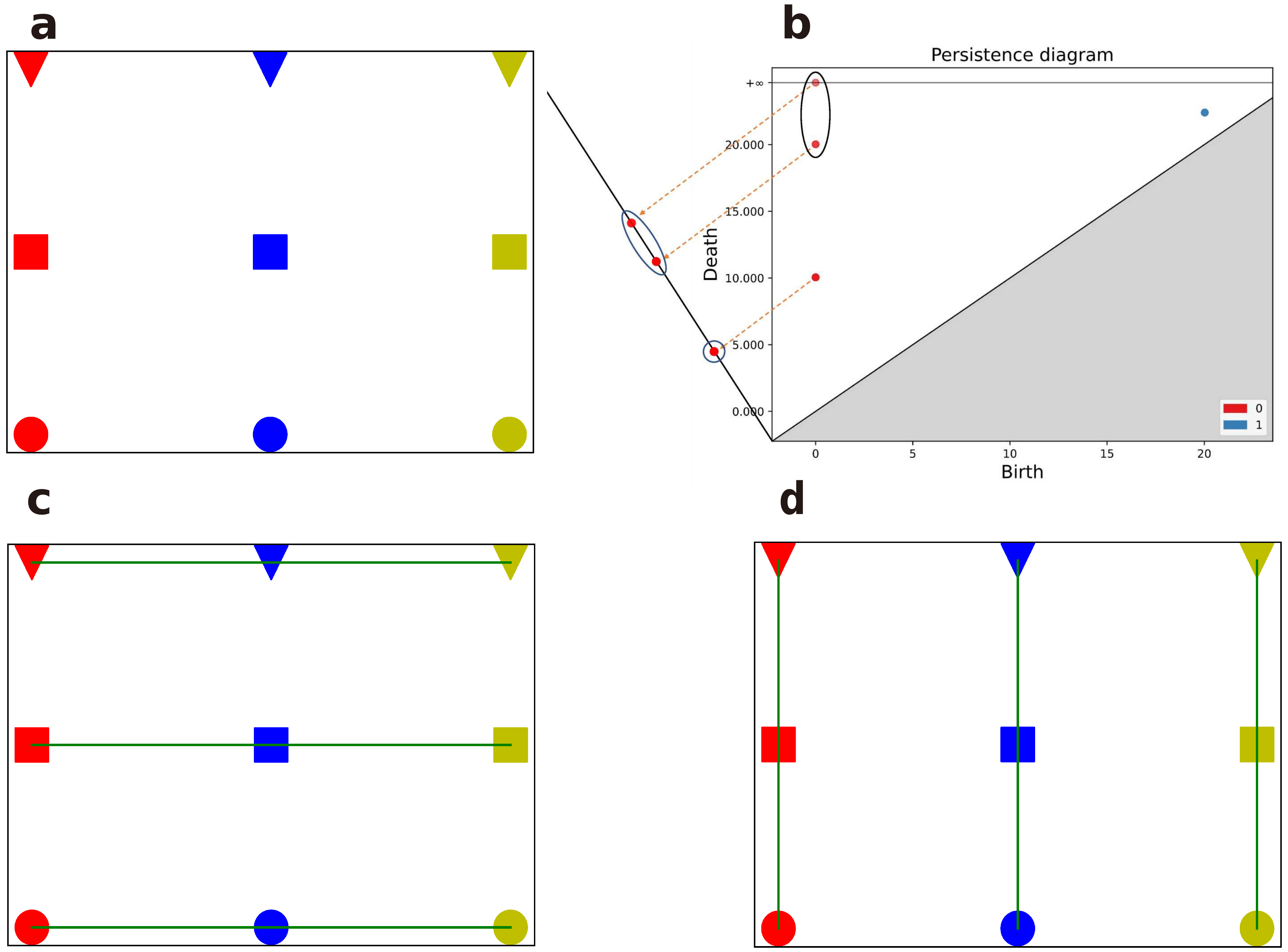}
 	\caption{Illustration of conflicts. 
 		a: An example of a conflict between different attributes of the similarity principle itself. 
 		b: The clustering result of the corresponding 0-PD with highlighted zero-dimensional significant points. Here two situations lead to the same 0-PD. 
 		c: The computation result when the shape feature is dominant. 
 		d: The computation result when the color feature is dominant.}
 	\label{fig:fig4}
 \end{figure*}
 
 \subsection*{Proximity principle}  
 The Gestalt proximity principle states that objects close or neighboring in space appear to be grouped together.
 The point set exemplifying the proximity principle is illustrated in Fig. \ref{fig:fig2} a.
 After clustering the points in the 0-PD (Fig. \ref{fig:fig2} b),
 	it is found that there are four points in the significant class,
 	corresponding to four connected components in the corresponding VR complex. 
 The computational result is the VR complex demonstrated in Fig. \ref{fig:fig2} c, 
 	which conforms to the Gestalt proximity principle,
 	i.e., the dots are perceived as four columns. 

 \subsection*{Closure principle}
 The Gestalt closure principle claims that though some shapes are not closed, 
 	our minds have a tendency to complete them,
 	filling in the gaps to perceive them as a whole. 
 Take the point set in Fig. \ref{fig:fig2} d as an example. 
 According to the Gestalt computational model, 
 	there is only one significant point in the 1-PD (Fig. \ref{fig:fig2} e), 
 	representing the hidden one-dimensional topological feature (i.e., loop) within the original point set (Fig. \ref{fig:fig2} d). 
 The VR complex with the significant topological feature, 
 	i.e., the computational result derived from Gestalt computational model,  
 	is illustrated in Fig. \ref{fig:fig2} f,
 	where the shape hidden in the original point set in Fig. \ref{fig:fig2} d is correctly closed.

 \subsection*{Good continuation principle}
 The Gestalt good continuation principle states that points that form straight lines or smooth curves 
 	when connected are perceived as belonging together, 
 	and these lines or curves tend to be seen as connected in the smoothest way possible.
 Now, we employ the Gestalt computational model to process the point set in Fig. \ref{fig:fig2} g.
 After clustering the points in the 1-PD (Fig. \ref{fig:fig2} h),
 	we get the threshold $\varepsilon_g = 1.2815$.
 The last step for the computation of good continuation, 
 	i.e., reconstructing the visual perceptual results,
 	is more complicated,
 	comparing with the Gestalt principles above. 
 First, we extract the 1-skeleton (which is composed of only 0-simplex and 1-simplex) \cite{edelsbrunner2022computational} of the VR complex
 	$VR(1.2815)$ (Fig. \ref{fig:fig2} i). 
 This 1-skeleton serves as a good approximation of the shape represented by the original point set. 
 Then, we select a starting point and an ending point in the 1-skeleton, 
 	corresponding to the two terminal nodes of an edge in the 1-skeleton in this example (Fig. \ref{fig:fig2} i). 
 Beginning from the starting point, 
 	we search for the next point along the direction with the smallest steering angle, 
 	continuing until the ending point is reached, thereby forming a branch of the shape (Fig. \ref{fig:fig2} j).
 In this way, each branch of the shape can be traced out (Fig. \ref{fig:fig2} j). 
 The shapes in Fig. \ref{fig:fig2} j calculated by the Gestalt computational model conform to the good continuation principle well. 

 \subsection*{Pragnanz principle}
 The Gestalt pragnanz principle, 
 	also known as simplicity principle, 
 	asserts that every stimulus will be perceived in the simplest possible manner.
 Specifically, during the process of perception, 
 	individuals tend to comprehend these global topological features in a straightforward manner, 
 	focusing on significant topological features and providing a simple understanding, 
 	while disregarding unimportant features. 
 For instance, one tends to perceive the shape of the five Olympic rings as the five circles (Fig. \ref{fig:fig3} a) 
 	rather than the nine sections (Fig. \ref{fig:fig3} b) \cite{goldstein1989sensation}.
 By the Gestalt computational model, 
 	considering the point cloud in Fig. \ref{fig:fig3} c that represents the shape of the five Olympic rings,
 	we cluster the points in the 1-PD (Fig. \ref{fig:fig3} d),
 	and then get five significant points in the significant class, 
 	corresponding to five significant features (the five circles, see Fig. \ref{fig:fig3} e).
 The VR complex $VR(\varepsilon_g)$ (Fig. \ref{fig:fig3} e) is the computational result, 
 	which keeps the significant topological features,
 	and discards the unimportant features, 
 	thus satisfying the pragnanz principle.

 \subsection*{Conflicts between different principles}
We can also deal with the conflicts between different Gestalt principles using the developed Gestalt computational model. 
In practice, conflicts between Gestalt principles, 
	such as the conflict between the proximity principle and the similarity principle, 
	as well as conflicts within the similarity principle itself due to different attributes, are common occurrences. 
	This presents an important research issue. 
Using persistent homology, 
	our model provides a quantitative method to explore these conflicts by selectively controlling a dominant principle or attribute.
An example of a conflict between different attributes of the similarity principle itself is depicted in Fig. \ref{fig:fig4}. 
	In the illustrated grid of points (Fig. \ref{fig:fig4} a), 
	each row shares the same shape, and each column shares the same color, 
	with two additional $z$-coordinate values added to each point,
	i.e., $z_{i,1}$ for color and $z_{i,2}$ for shape. 
	Consequently, each planar point corresponds to a 4-dimensional coordinate, where each point in the point set~(\ref{eq:pt_set}) is located at the barycenter of each shape.
By adjusting the salience of these attributes, 
	we can control which of the two will be the dominant attribute. 
	For example, if we assign a greater difference value to the shape than to the color, 
	then upon computing and clustering the 0-PD and selecting parameter $\varepsilon_g$, 
	we obtain the VR complex $VR(\varepsilon_g)$. 
	This VR complex illustrates that the points will be classified according to the shape attribute (Fig. \ref{fig:fig4} c). 
Conversely, if we assign a greater difference value to the color than to the shape, 
	again by computing and clustering the 0-PD and selecting $VR(\varepsilon_g)$, 
	we can see that the points will be classified according to the color attribute (Fig. \ref{fig:fig4} d). 
This suggests that the Gestalt computational model, 
	rooted in persistent homology, 
	can effectively function as a versatile quantitative tool for investigating conflicts between Gestalt principles.

\begin{figure*}[ht]
	\centering
	\includegraphics[width=1.0\linewidth]{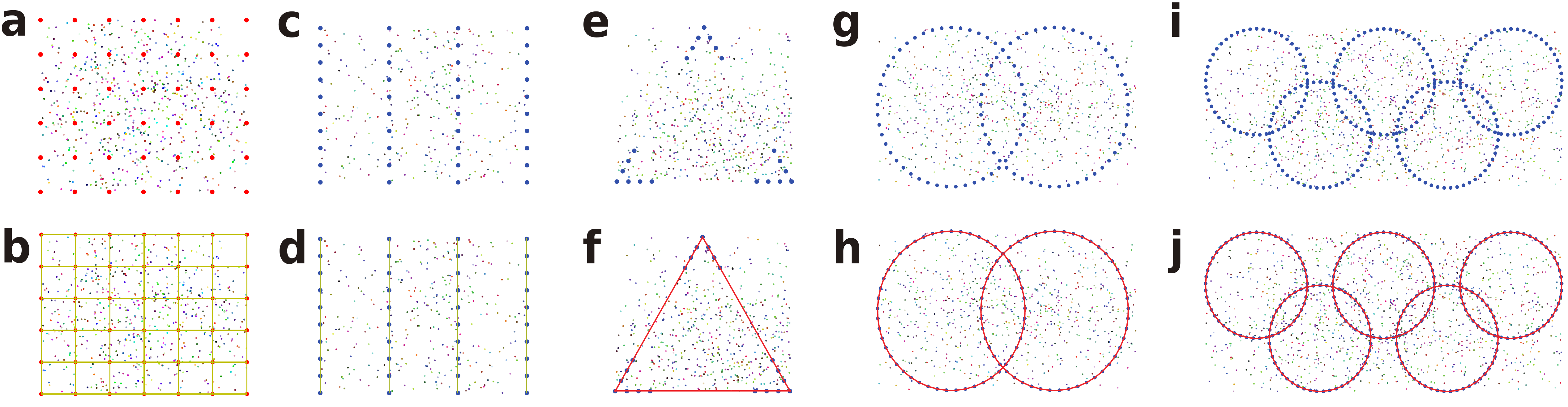}
	\caption{Point sets with clutter points.
		a, c, e, g, i: Point sets we have used in previous sections, with additional random clutter points adhering to a Gaussian distribution and random colors.
		b, d, f, h, j: Computational results of our model. To improve clarity, simplices formed by clutter points are omitted. Results aligning with Gestalt theory can still be reconstructed.
	}
	\label{fig:so}
\end{figure*}
\begin{figure*}[ht]
	\centering
	\includegraphics[width=1.0\linewidth]{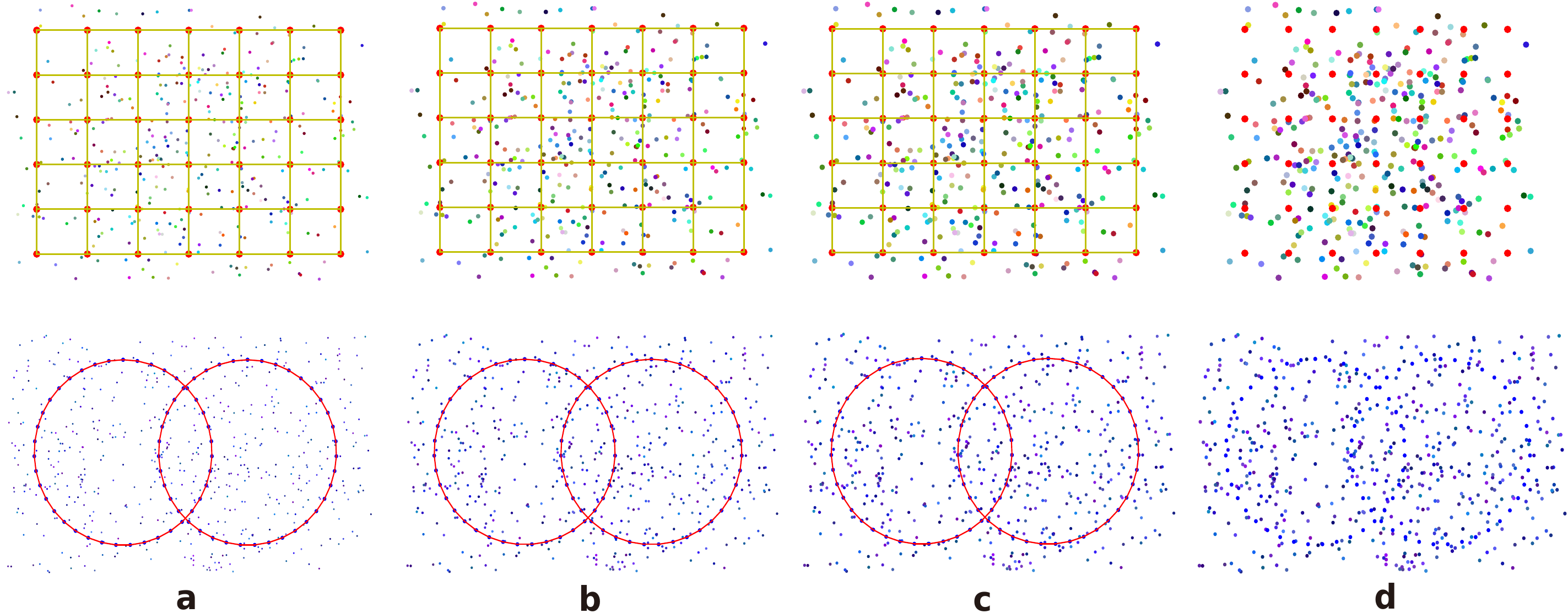}
	\caption{From a to d, the size of clutter points increases, and the Gestalt principles are destroyed when the size of clutter points is close to the original points (d). 
	(a, b, c) The correct shapes are generated by the Gestalt computational model. 
	(d) The Gestalt computational model  fails due to the destruction of the topological features formed by Gestalt principles. 
	}
	\label{fig:noisecomp}
\end{figure*}
\subsection*{Computational results on point sets with clutter points}
We now conduct tests on more complex examples to demonstrate the effectiveness of our computational model. In Fig. \ref{fig:so}, random clutter points are added to the point sets used in previous sections (Fig. \ref{fig:so} a, c, e, g and i). These clutter points adhere to a Gaussian distribution and possess randomly selected colors. Because the clutter points are smaller in size compared to the original point set, it is still feasible to reconstruct a visual perception result from the original point set that conforms to the Gestalt principles. Employing our computational model, we assign each point $z_{i,1}$ the normalized hue value  \cite{ibraheem2012understanding} of its color within the range $[0,1]$, 
	while we assign $z_{i,2}$ as the area of each point, where the larger points are ten times the area of the smaller points (here the area of the point becomes the main attribute rather than the color).
The results demonstrate that despite the presence of clutter points causing a messy background, the Gestalt computational model can still produce outcomes that adhere to visual perception, since the essential Gestalt features of the initial point sets remain intact.

In contrast, if the size of clutter points is close to that of the original points, it can be challenging for even humans to differentiate between them, since the Gestalt principles presented in the shapes are destroyed. For instance, in Fig. \ref{fig:noisecomp}, it is illustrated that as the size of clutter points increases, the Gestalt principles are compromised. 
When the size of clutter points is close to that of the original points, it results in the destruction of the topological features formed by Gestalt principles (Fig.~\ref{fig:noisecomp} d).
In these cases, as the Gestalt principles presented in the shapes are disrupted, our computational model no longer produces the previous computation results. 
These results also align with human visual perception.

\begin{figure*}[ht]
	\centering
	\includegraphics[width=1.0\linewidth]{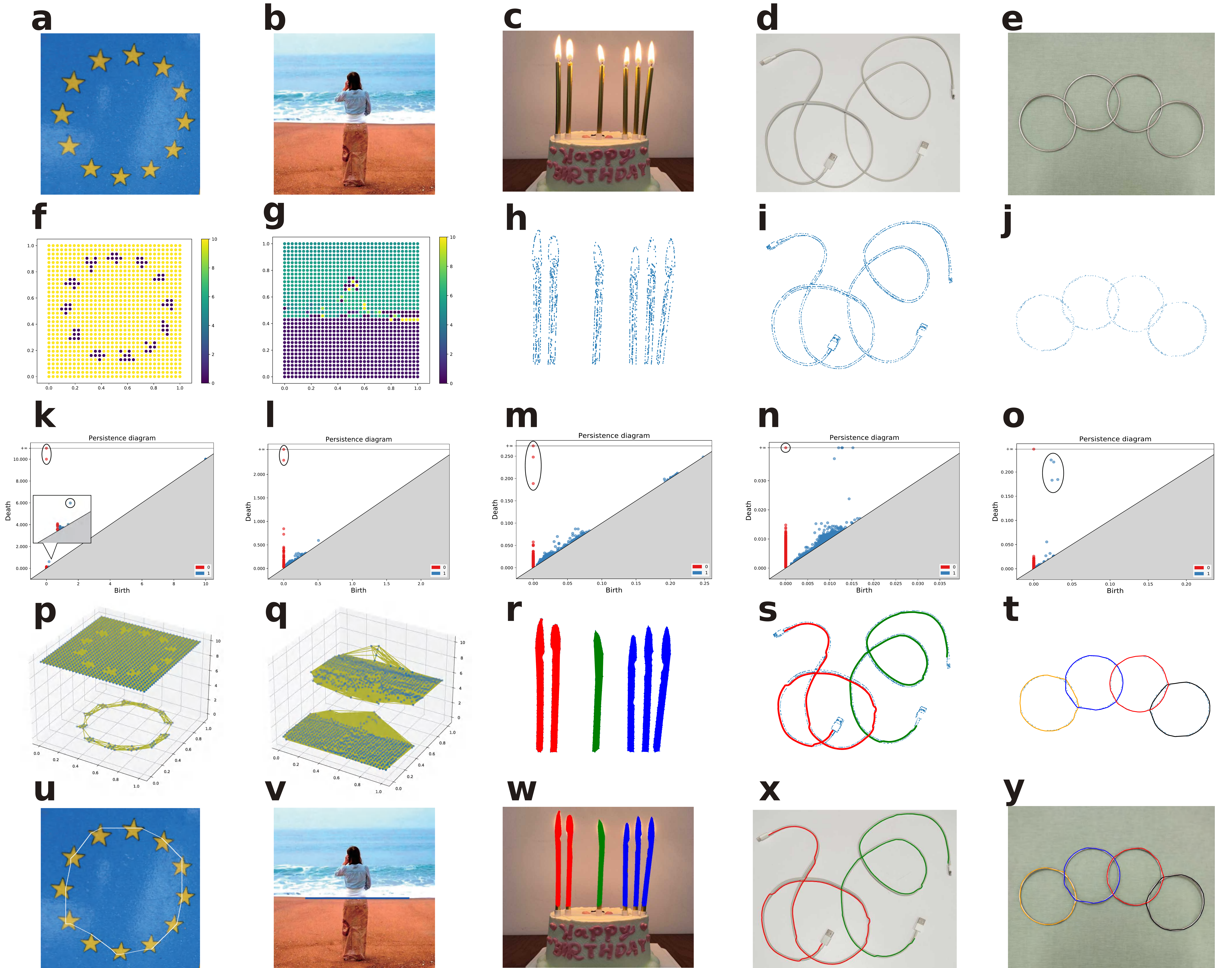}
	\caption{Real image examples. a--e: Five real images.  
		f--j: Point sets obtained from the original images through image processing methods such as color recognition and edge detection, respectively.  
		k--o: Clustering results of PDs, highlighting only significant points in each PD.  
		p--t: Computational results of the five images using our model, respectively.  
		u--y: Visualization of the reconstructed perceptual results on the original images.  
	} 
	\label{fig:images}
\end{figure*}

\subsection*{Computational results on the images}
Here, we present some results obtained by applying the computational model to some real images, exemplifying various Gestalt principles. It is important to note that before applying our Gestalt computational model, image processing methods such as color recognition, edge detection, etc., must be employed to prepare the image for compatibility with the computational model.

The first image is that of the European Union flag (Fig. ~\ref{fig:images} a) obtained from the Penn-State symmetry recognition competition data \cite{liu2013symmetry}. In this image, the similarity of colors leads to grouping, with the twelve stars tending to be perceived as forming a closed circle, demonstrating Gestalt similarity and closure principles. To utilize the Gestalt computational model, an additional coordinate, $z_{i,1}$, is required to represent the color information. In this case, we assign $z_{i,1}$ to each pixel based on their hue value: points with hues closer to yellow are assigned $z_{i,1}=0$, while points with hues closer to blue are assigned $z_{i,1}=10$. For computational convenience, we densely and uniformly sample the pixel points on the image and normalize the location coordinates of the sampled pixel points within the unit square $[0,1]\times [0,1]$ (Fig. \ref{fig:images} f).
By computing and clustering the 0-PD and 1-PD respectively (Fig.~ \ref{fig:images} k), two zero-dimensional significant points and one one-dimensional significant point are identified. Upon visualizing the results in a 3D view (Fig.~ \ref{fig:images} p), two connected components are distinguished by color, with a hidden loop apparent among the twelve yellow stars, aligning with the Gestalt similarity and closure principles. Utilizing our computational models, we can extract the hidden loop, as illustrated in Fig. ~\ref{fig:images} u.

The second image, \textit{Waves} (Fig.~ \ref{fig:images} b), by Wilma Hurskainen, captures the precise moment when the front of the white water aligned with the white area on the woman’s clothing. In this image, color similarity leads to grouping, particularly different colored areas of the woman’s clothing being perceptually grouped with the same colors in the scene. Additionally, the front edge of the water demonstrates good continuation as it crosses the woman’s clothing. Thus, this image exemplifies Gestalt similarity and good continuation principles \cite{goldstein1989sensation}. 
Given the richer colors in this image, we directly utilize the hue value of each pixel as an additional $z_{i,1}$ coordinate value. For computational convenience, we linearly transform the hue values to the range $[0,10]$, and continue to densely and uniformly sample the pixel points on the image, normalizing the location coordinates of the sampled pixel points within the unit square $[0,1]\times [0,1]$ (Fig. \ref{fig:images} g).
By computing and clustering the 0-PD, two zero-dimensional significant points are identified (Fig. \ref{fig:images} l), depicting two connected components grouped by color (Fig. \ref{fig:images} q).
Based on our calculations, selecting two pixel points on the boundary of one main part of grouping components (here the two points are the two terminal nodes of the blue line segment in Fig. \ref{fig:images} v), and employing the computation of the good continuation principle in our computational model, we derive the blue line segment in Fig. \ref{fig:images} v, which precisely divides the woman's clothing into two parts based on color. Thus, the results align with the Gestalt similarity and good continuation principles.

The third image shows five candles (Fig. \ref{fig:images} c), but the two on the left are closer together and would be considered a group, the three on the right are closer together and would be considered another group, and the separate one in the center would be considered a separate group. Thus, these candles would be perceived as three groups in total. This image illustrates the Gestalt proximity principle very well \cite{goldstein1989sensation}. Now we use our model to compute this principle. Initially, the Canny edge detection method \cite{canny1986computational} is applied to sample pixels approximating the five candles, followed by normalizing the location coordinates into the unit square $[0,1]\times [0,1]$ (Fig. \ref{fig:images} h). Here no extra coordinates are required. Subsequently, we compute and cluster the corresponding 0-PD, obtaining three significant points (Fig. \ref{fig:images} m). By selecting $\varepsilon_g$ and filling each candle with the produced 2-simplices, then overlaying the results on the image where each group is indicated by one color, we obtain visual perceptual results that align with the Gestalt proximity principle, as shown in Fig. \ref{fig:images} r and w.

The fourth image displays interleaving wires (Fig. \ref{fig:images} d), demonstrating the Gestalt good continuation principle, allowing the perception of two wires \cite{goldstein1989sensation}. We now utilize our model to compute these perceptual results. Initially, the Canny edge detection method is employed to sample pixels approximating the outline of the two wires, followed by normalizing the location coordinates into the unit square $[0,1]\times [0,1]$ (Fig. \ref{fig:images} i). Here no extra coordinates are required. Subsequently, we compute and cluster the 0-PD (Fig. \ref{fig:images} n), determining $\varepsilon_g$ such that the 1-skeleton $VR(\varepsilon_g)$ can approximate the wire outlines. For each wire, we select a starting point and an ending point, utilizing the good continuation principle to extract two curves. The curves representing the results are displayed in Fig. \ref{fig:images} s and overlaid on the image, as depicted in Fig. \ref{fig:images} x. Due to the noise generated by the edge detection process, the computational results slightly deviate from the actual wire positions in some areas of the image. However, overall, they still approximate the shape of the wires and adhere to the Gestalt good continuation principle.

The last image depicts the linking rings (Fig. \ref{fig:images} e), a common magical prop. In accordance with the Gestalt pragnanz principle, audiences will perceive it as four separate rings during a magic performance. Consequently, a magician crafts scenes that intentionally contradict this principle of visual perception to achieve the desired effect during their performance. Using the Gestalt computational model, we can compute these perceptual results. We first employ the Canny edge detection method to sample pixels approximating the outline of the rings, followed by normalizing the location coordinates into the unit square $[0,1]\times [0,1]$ (Fig. \ref{fig:images} j). Here no extra coordinates are required. Then we compute and cluster the 1-PD, resulting in four significant points (Fig. \ref{fig:images} o). Following the selection of $\varepsilon_g$ and derivation of the 1-skeleton $VR(\varepsilon_g)$, the Gestalt good continuation principle is applied to extract four rings (Fig. \ref{fig:images} t and y), corresponding to the four significant points in the 1-PD. Hence, the computational results conform to Gestalt pragnanz principle.

\section*{Discussion}
We have developed a computational model for Gestalt theory using persistent homology, 
	which integrates Gestalt theory from cognitive psychology with persistent homology in computational topology for the first time. This provides a clear description of the relationship between the global and local aspects in Gestalt theory.
	Additionally, the capacity of persistent homology to extract global topological features makes it a valuable tool for calculating Gestalt theory. 
Furthermore, our model has the potential to advance research in the quantitative aspects of Gestalt theory and to serve as a novel method for investigating human visual perception in computational psychology.

\section*{Methods}
\subsection*{Vietoris--Rips complex}
The $n$-simplex is defined as the convex hull formed by $n+1$ affine independent points $\left\lbrace u_0, u_1, \ldots, u_n \right\rbrace$ in Euclidean space $\mathbb{R}^N$, denoted as $[u_0, u_1, \ldots, u_n]$. An $n$-simplex can be represented by various geometric models, such as a vertex (0-simplex), a line segment (1-simplex), a triangle (2-simplex), and a tetrahedron (3-simplex). A (abstract) simplicial complex $K$ is a collection of simplices that satisfies the following properties: every face of a simplex in $K$ is also in $K$, and the intersection of any two simplices in $K$ is a face of both of them \cite{munkres2018elements}. Moreover, we define the dimension of a simplicial complex $K$ as the maximum dimension among all the simplices in $K$.

The Vietoris--Rips (VR) Complex \cite{edelsbrunner2022computational} is a type of simplicial complex. Its construction is defined by the following rules: for any $\varepsilon>0$, a finite subset $\{x_0, x_1, \ldots, x_n\} \subseteq X$ of the space $\mathbb{R}^N$ forms a simplex $[x_0, x_1, \ldots, x_n]$ if and only if the distance between any pair of points $x_i$ and $x_j$ satisfies $d(x_i, x_j) \leq \varepsilon$. The collection of all simplices generated by the point cloud $X$ that satisfy the aforementioned conditions constitutes the VR complex. It is worth noting that alternative definitions may employ a bound of $2\varepsilon$ instead of $\varepsilon$ (i.e. $d(x_i, x_j) \leq 2\varepsilon$), resulting in the same combinatorial object but with filtration parameters that are half. Although this choice affects the filtration parameters, it does not alter the underlying combinatorial structure of the complex. Consequently, the extraction results of topological features are not affected.

\begin{figure*}[ht]
	\centering
	\includegraphics[width=0.8\linewidth]{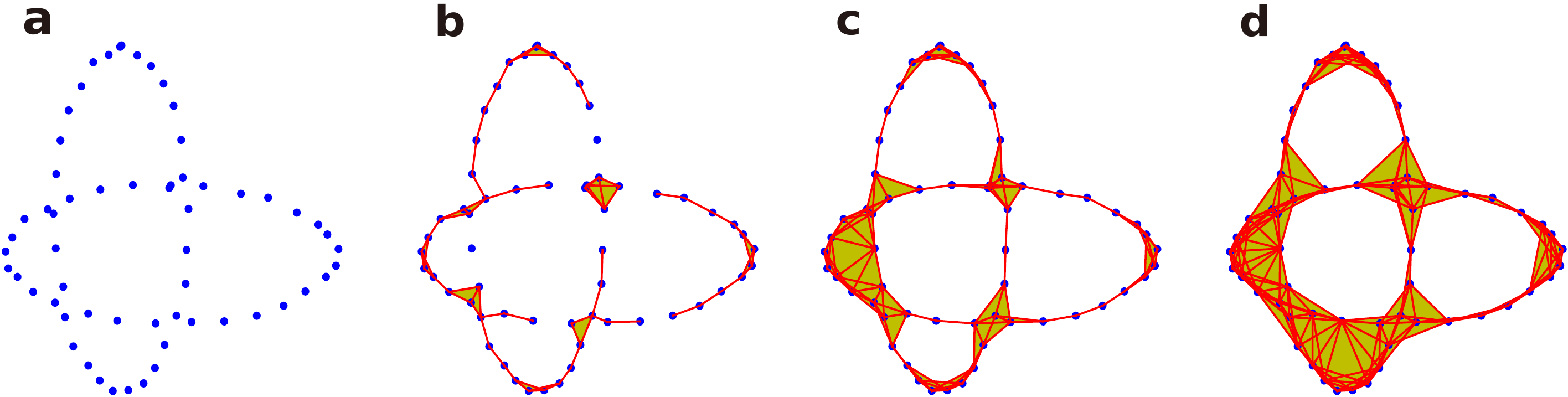}
	\caption{An example of VR filtration with parameter $\varepsilon$. a, b, c, d corresponds to $\varepsilon=0.0,1.0,1.5,2.0$ respectively. Here only simplices with dimension less than and equal to two are depicted. }	\label{fig:vr}
\end{figure*}

\subsection*{Persistent homology}
 Let $K$ be a simplicial complex and $n$ its dimension. An $n$-chain is a sum of $n$-simplices in $K$, denoted by $c=\sum a_i\sigma_i$, where $\sigma_i$ represents an $n$-simplex, and $a_i$ is its coefficient belonging to an Abelian group, typically $\mathbb{Z}$ or $\mathbb{Z}_2$ (in our work we use $\mathbb{Z}_2$ coefficients). With the addition operation, the $n$-chains form the $n$-chain group, denoted by $C_n(K)$.

 The boundary for a $n$-simplex $\sigma=[x_0,x_1,\ldots,x_n]$ is given by
 $$
 \partial_{n}\sigma=\sum_{j=0}^{n}(-1)^{j}[x_{0},\ldots,\hat{x}_{j},\ldots,x_{n}],
 $$
 where $\hat{x}_j$ denotes the omission of $x_j$. Let $Z_n(K)=\operatorname{Ker}(\partial_n)$ and $B_n(K)=\operatorname{Im}(\partial_{n+1})$. The $n$-th homology group of $K$ is defined as the quotient group $H_{n}(K)=Z_{n}(K)/B_{n}(K)$.

 Persistent homology offers a multi-scale description of homology using filtration. 
 Given a point cloud $X$ and a series of parameters $0=\varepsilon_0<\varepsilon_1<\cdots<\varepsilon_m$, the nested VR complex sequence	
 	$$
 	VR(X,\varepsilon_{0})\subset VR(X,\varepsilon_{1})\subset\cdots\subset VR(X,\varepsilon_{m})
 	$$
 is referred to as a VR filtration \cite{edelsbrunner2022computational}. Without confusing notation, we will also write $VR(X,\varepsilon)$ simply as $VR(\varepsilon)$.  Fig. \ref{fig:vr} shows an example of a VR filtration. If we regard each $\varepsilon$ as one moment, then at each moment in the filtration there is a different VR complex, which is called the state corresponding to $\varepsilon$ in the filtration.

 If we denote each $VR(X,\varepsilon_i)$ as $K_i \:(0\le i \le m)$, then for every $i\leq j$ we have an inclusion map from $K_i$ to $K_j$ and therefore an induced homomorphism $f_p^{i,j}: H_p(K_i)\to H_p(K_j)$ for each dimension $p$. The $p$-th persistent homology groups \cite{edelsbrunner2022computational} are defined as the images of the homomorphisms induced by inclusion:
 $$H_p^{i,j}=\operatorname{Im}f_p^{i,j},\: \forall  0\leq i\leq j\leq m.$$ 
 The corresponding $p$-th persistent Betti numbers are the ranks of these groups: $\beta_{p}^{i,j}=\operatorname{rank}H_{p}^{i,j}$.

 The standard method for computing the persistent homology of a filtration is the reduced matrix method \cite{edelsbrunner2022computational}, and there are corresponding tools for computing persistent homology within many software packages and libraries. As the parameter $\varepsilon$ increases, we can observe the birth time and death time of topological features (i.e. the representations of generators of persistent homology groups, which are also called cycles) in different dimensions. One of the most common tools for visualizing persistent homology is the persistence diagram (PD) \cite{edelsbrunner2022computational}, as shown in Fig. \ref{fig:PD} a, which can be calculated using tools such as the GUDHI Python module \cite{maria2014gudhi}. The set of points that records the birth time and death time of the $n$-cycles are denoted as the $n$-PD. Each point in an $n$-PD takes the form $(b_i, d_i)$, representing an $n$-cycle and capturing its birth time $b_i$ and death time $d_i$. It is clear that all points in a PD are located above the diagonal $y=x$. And the value $|d_i - b_i|$ is denoted as the persistence of this cycle. Moreover, a point in a persistence diagram is termed a multi-point if at least one other point exists at its location.

\subsection*{Clustering of points in a persistence diagram}
 We can identify the significant topological features of the underlying space from the persistence diagrams. To achieve this, it is necessary to locate the points in the $n$-PD that represent $n$-dimensional significant topological features. An effective approach involves clustering the points in the $n$-PD and extracting the class that exhibits the greater persistence.

\begin{figure*}[ht]
	\centering
	\includegraphics[width=0.7\linewidth]{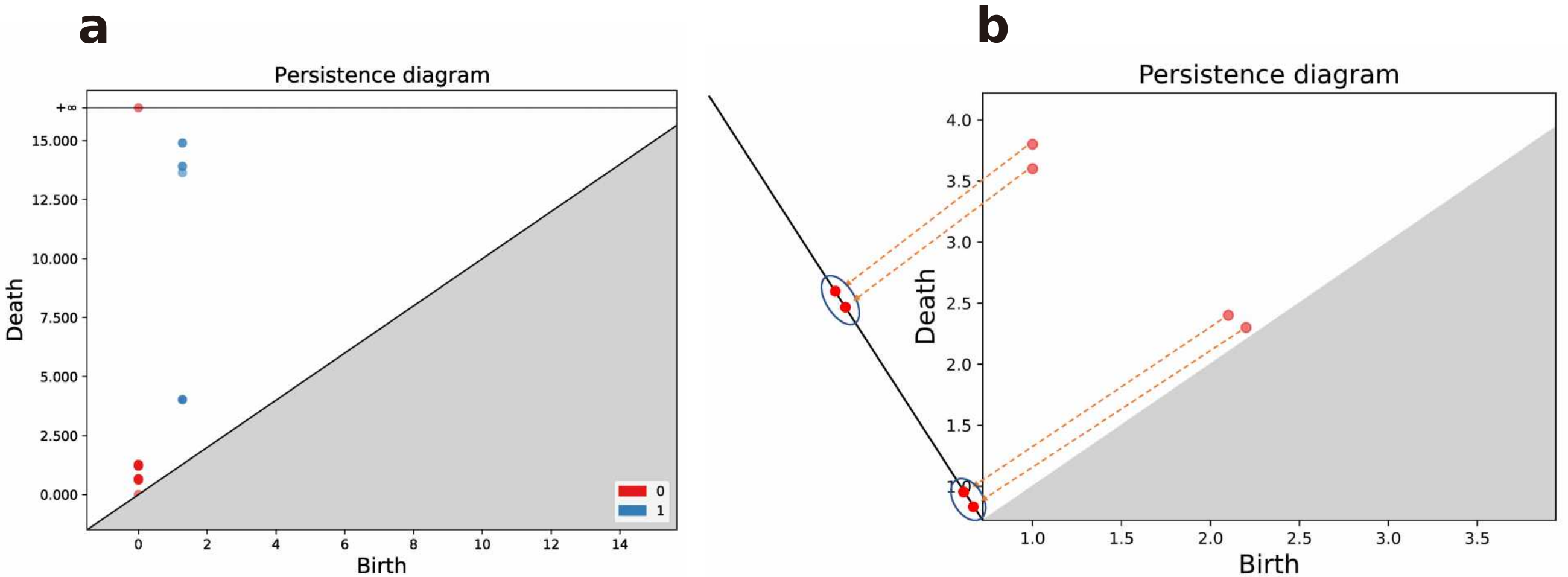}
	\caption{Illustration of persistence diagram. 
		a: An example of persistence diagram. As the label shows, the red points in the persistence diagram represent the zero-dimensional persistence diagram, and the blue points represent the one-dimensional persistence diagram. 
		b: The process of clustering points in the given $n$-PD based on persistence.}	
	\label{fig:PD}
\end{figure*}

 To complete the clustering, initially we project the points in the given $n$-PD onto the line $y = -x$, which is perpendicular to the diagonal $y=x$ in the PD. It becomes apparent that the distances from the projected points to the origin $(0, 0)$ can effectively represent the persistence of the relevant points in the PD. If a point has a death time of $+\infty$, we can either skip it and directly classify it as a significant point, or replace its $+\infty$ coordinate with a suitably large value that will not affect the clustering results during the clustering process. For example, we can replace it with a value that is appropriately larger than the second largest death time among the other points. Subsequently, the projected points on $y = -x$ can be classified into two classes using a clustering algorithm, such as the $k$-means algorithm ($k=2$). Fig. \ref{fig:PD} b illustrates the process of clustering a PD. The points within the class demonstrating greater persistence correspond to significant topological features, referred to as significant points, while points within the other class correspond to noise features, referred to as noise points. If the PD contains only one point with positive persistence, we consider this point to be significant, indicating the existence of only the significant class. In our method, we primarily focus on the significant points in the PDs to illustrate the topological understanding of visual perception. Therefore, we will highlight these significant points on the obtained PDs.

\normalem
%\bibliographystyle{unsrt}
%\bibliography{bib_G}

\section*{Data availability}
The authors declare that the data supporting the findings of this study are available within the paper and supplementary information files.

\subsection*{Acknowledgments}
This work is supported by the National Natural Science Foundation of China under Grant Nos. 62272406 and 61932018. 

\subsection*{Author contributions statement}
Y.C. and H.L. carried out the theoretical analysis and wrote the manuscript. 
Y.C., H.L. and J.Y. devised the experiments.
Y.C. and H.L implemented the code of the computational model.
All authors participated in editing the manuscript.

\subsection*{Competing interests}
The authors declare no competing interests.
  
\subsection*{Additional information}
Supplementary Information is available for this paper.

%\end{multicols}
\end{document}